\documentclass[preprint,showpacs,floatfix,amsmath,amssymb]{revtex4}
\usepackage{graphicx}% Include figure files
\usepackage{dcolumn}% Align table columns on decimal point
\usepackage{bm}% bold math
\usepackage{amssymb}
\usepackage{graphicx}
\usepackage{amsmath}
\usepackage{xspace}
\begin{document}

\title{Superconductivity at 33 K in ``111" single crystals at ambient pressure}
\author{T.-L. Xia, J. B. He, D. M. Wang, and G. F. Chen}

\email{genfuchen@ruc.edu.cn}

\affiliation{Department of Physics, Renmin University of China,
Beijing 100872, P. R. China}
%\affiliation{$^{2}$Beijing National Laboratory for Condensed Matter
%Physics, Institute of Physics, Chinese Academy of Sciences, Beijing
%100190, P. R. China}
\date{\today}
\begin{abstract}
We have successfully grown single crystalline NaFeAs with cobalt
or phosphor doping. Bulk superconductivity occurs in
NaFe$_{0.95}$Co$_{0.05}$As at 19 K while much higher transition
temperature is observed in NaFeAs$_{0.8}$P$_{0.2}$, in which the
Tc of 33 K is even higher than the highest value realized by
applying pressure in NaFeAs and all other isoelectronic element
substituted samples. We discuss these behaviours by comparison
with 122 and 1111 systems. We hope that our findings will help
 improve our understanding of iron-based superconductivity.
\end{abstract}


\maketitle

Since the discovery of superconductivity at 26 K in iron-based
oxypnictide LaFeAsO$_{1-x}$F$_{x}$,\cite{kamihara} doping at
different sites in the corresponding parent compound LnFeAsO
(Ln=Lanthanide) with different elements has been extensively
attempted to get new superconductors, including isoelectronic
substitution of smaller lanthanide ions for bigger ones leading
the Tc in the iron-based superconductors firstly to be higher than
40 K,\cite{gfchen,xhchen} oxygen deficiency in the 1111 system to
give higher transition temperature up to 55 K,\cite{zaren} and the
carriers doping with different types at different sites inducing
superconductivity.\cite{hhwen,cwang,sefat} Soon after the
discovery of above so called 1111 system with ZrCuSiAs-type
structure, AFe$_{2}$As$_{2}$ (A=Ca, Sr, Ba, Eu) with
ThCr$_{2}$Si$_{2}$-type structure (122 system) were found to
superconduct once the carriers are appropriately
introduced\cite{rotter,gfchen122,sasmal,ni,sefat2,jasper} or the
pressure was applied.\cite{torikachvili,torikachvili2} K doping at
A site,\cite{rotter,gfchen122,sasmal,ni} P doping at As
site\cite{zren} and Co doping at Fe site\cite{sefat2,jasper} all
lead to superconductors, in which the highest Tc is 38 K in
Ba$_{1-x}$K$_{x}$FeAs.\cite{rotter}

As for 111 system, the investigation on LiFeAs presents
superconductivity without any hint on structural or magnetic
orders.\cite{xcwang, pitcher, tapp} The same situation holds for
NaFeAs before we present the high quality data on single
crystalline NaFeAs, where three successive transitions
corresponding to structural, magnetic and superconducting
transitions have been observed in the resistivity and magnetic
susceptibility curves\cite{gfchen111} and later confirmed by
neutron scattering and muon-spin rotation
results.\cite{slli,parker} The neutron scattering results indicate
that the spin within ab plane is ferromagnetically aligned along b
axis and antiferromagnetically along a axis, while in c directions
it is antiferromagnetically coupled. Up to now, there is no
reports on the study of magnetic structure in LiFeAs, but Li
\emph{et al.} did the first-principles calculations and predicted
the stripe antiferromagnetic ground state with easy axis of
magnetization along the b direction of the magnetic unit
cell.\cite{zli} Thus, 111 system is believed to be the same as 122
and 1111 systems in their magnetic structures. But if we take into
account the position of Na or Li ions with respect to the FeAs
block, we see difference between 111 system and 122 or 1111
system. Li or Na takes the position of Fe(2) in the compound of
Fe$_{2}$As (PbFCl or Cu2Sb structure) and it is assumed to be
possible that Li or Na directly interacts with the 5 neighboring
As. The delicate difference in their crystal structures may be one
possible origin for their slightly different electronic properties
between LiFeAs (NaFeAs) and 1111 or 122 systems. To elucidate this
problem, NaFeAs is doped with Co or P to see if the results can
provide us some hints on the difference or common features between
their properties and other doped 122 or 1111 system and further
supply additional data to understand the mechanism of
superconductivity in Fe-Pnictide superconductors.

\begin{figure}[t]
\includegraphics[width=18 cm]{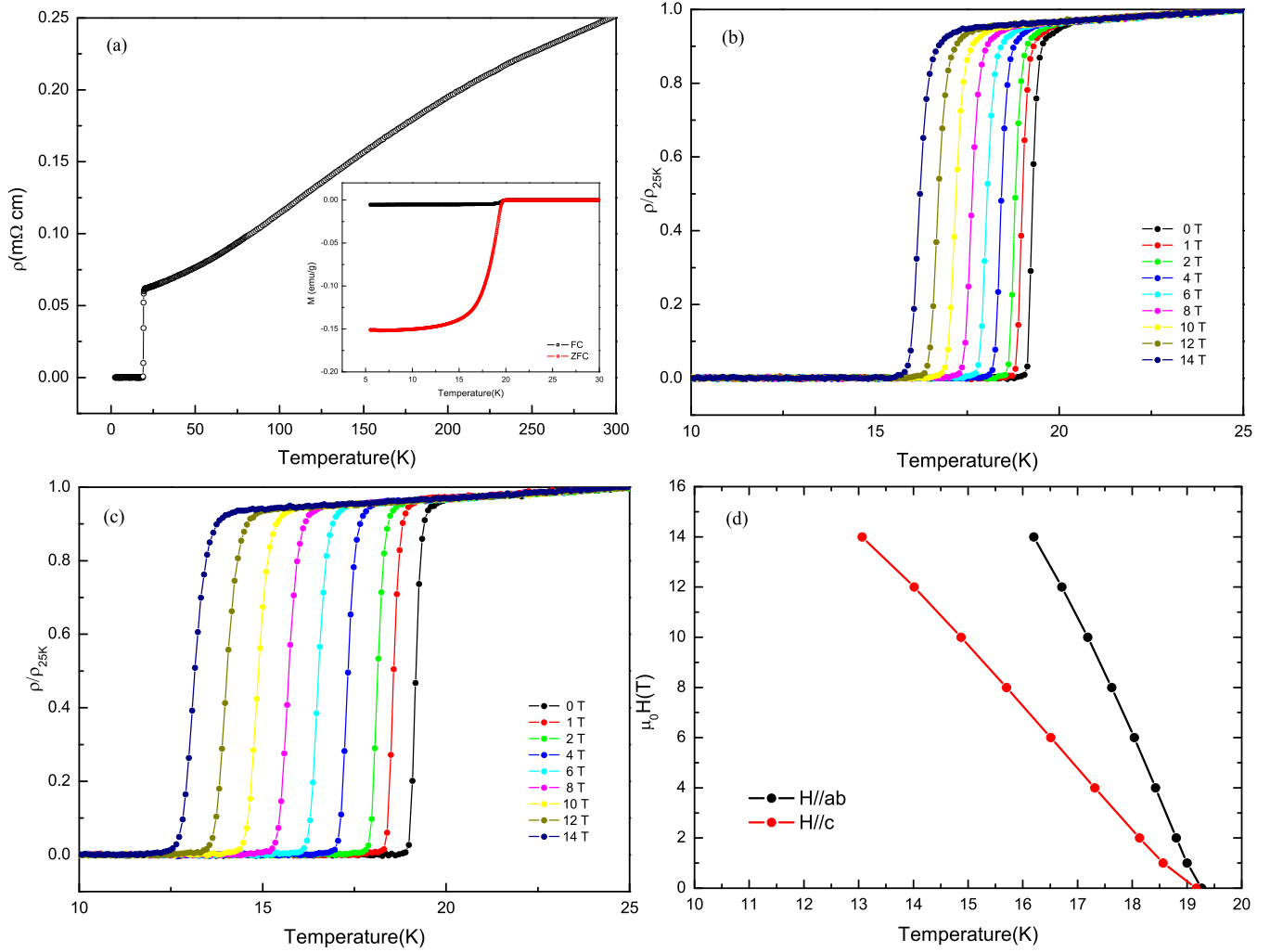}
\caption{(Color online) (a) Temperature dependence of the
electrical resistivity for a NaFe$_{0.95}$Co$_{0.05}$As single
crystal in zero field up to 300 K. The inset shows the temperature
dependent magnetization data under ZFC and FC conditions with H=10
Oe. (b) Temperature dependence of the resistivity normalized to
resistivity at 25 K at fields up to 14 T with H$\|$ab. (c)
Temperature dependence of the resistivity normalized to
resistivity at 25 K at fields up to 14 T with H$\|$c. (d) The
upper critical field of NaFe$_{0.95}$Co$_{0.05}$As single crystal
for H$\|$ab and H$\|$c.}
\end{figure}
 Both kind of samples were grown employing self-flux method. NaAs,
FeAs, Fe$_{2}$As, CoAs, FeP, and Fe$_{2}$P were firstly
synthesized as precursors in the same way as reported
before\cite{gfchen111}. Mixtures of above raw materials with
nominal compositions NaFe$_{0.95}$Co$_{0.05}$As and
NaFeAs$_{0.8}$P$_{0.2}$ were loaded into alumina tubes and then
sealed into Ta tubes with Ar under the pressure of 1.5 atom., then
the sealed Ta tubes were vacuum sealed into quartz tubes. The
samples were put in box furnace and heated to 1170 $^{0}$C slowly
and held there for a few hours before the temperature was
decreased very slowly. The grown crystals have the form of
platelets with shiny facets. The dimensions of the crystals are
not so large as 122 crystals. The maximum size of the grown and
mechanically cleaved samples are $\sim$5$\times$5$\times$0.2
mm$^{3}$. Their electronic and magnetic properties were measured
on a Quantum Design physical property measurement system (PPMS)
with the VSM option provided.

Figure 1(a) shows the temperature dependence of resistivity for
NaFe$_{0.95}$Co$_{0.05}$As crystal between 2.2 K and 300 K. No
other anomalies are observed except for the superconducting
transition at 19 K (Here Tc is defined as the temperature where
zero resistivity is achieved), which means no transitions related
to magnetic or structural orders exist as found in
Na$_{1-\delta}$FeAs single crystals.\cite{gfchen111} The result is
consistent with previous reports on Co doped NaFeAs
polycrystalline samples,\cite{parker2} where the magnetic and
structural transitions are absent when the carriers like Co or Ni
are introduced into the system and the bulk superconductivity is
established. The inset of Fig. 1(a) shows the temperature
dependence of the magnetization in the single crystal of
NaFe$_{0.95}$Co$_{0.05}$As under the conditions of zero field
cooling (ZFC) and field cooling (FC) at H=10 Oe. The perfect
diamagnetism indicates the bulk superconductivity. Taking into
account the Tcs in other Co doped system like 22 K in
BaFe$_{2-x}$Co$_{x}$As$_{2}$, 20 K in
SrFe$_{2-x}$Co$_{x}$As$_{2}$, 18-19 K in
CaFe$_{2-x}$Co$_{x}$As$_{2}$ and 10-24 K in
LnFe$_{1-x}$Co$_{x}$AsO (Ln=La, Ce, Nd, Sm \emph{etc.}), the
transition temperature of 19 K is at the same level and consistent
with previous reports on polycrystalline Co-doped
NaFeAs.\cite{parker2} Later in this paper we find P-doped NaFeAs
presents much higher Tc compared with above values and it is even
higher than all the Tcs realized through Co or P doping in their
parent compounds in 122 or 1111 system.

Figure1 (b) and (c) show the behavior of resistivity of
NaFe$_{0.95}$Co$_{0.05}$As in external magnetic field up to 14 T.
In Fig. 1(b), the applied field is within the ab plane ($H\| ab$)
while in Fig. 1(c) the applied field is parallel to the c axis
($H\| c$). We see the superconducting transition is suppressed in
both conditions but the effect of magnetic field is much larger
when the field is applied along the c axis of the single crystals
in stead of within the ab plane. The corresponding upper critical
field $H_{c2}$ as a function of temperature obtained from a
determination of the midpoint of the resistive transition is
plotted in Fig. 1(d). The curves are steep with slopes
$-dH_{c2}/dT|_{T_{c}}$=4.7 T/K for $H\|ab$ and
$-dH_{c2}/dT|_{T_{c}}$=2.4 T/K for $H\|c$. According to the
Werthamer-Helfand-Hohenberg (WHH) formula\cite{WHH}
$H_{c2}(0)=-0.69(dH_{c2}/dT)T_{c}$ and taking 19 K as Tc, the
upper critical fields are estimated to be $H_{c2}^{ab}$=62 T and
$H_{c2}^{c}$=31 T. The ratio of $H_{c2}^{ab}/H_{c2}^{c}$ is about
2 which means the anisotropy is not so large even the upper
critical fields are high.

\begin{figure}[t]
\includegraphics[width=18 cm]{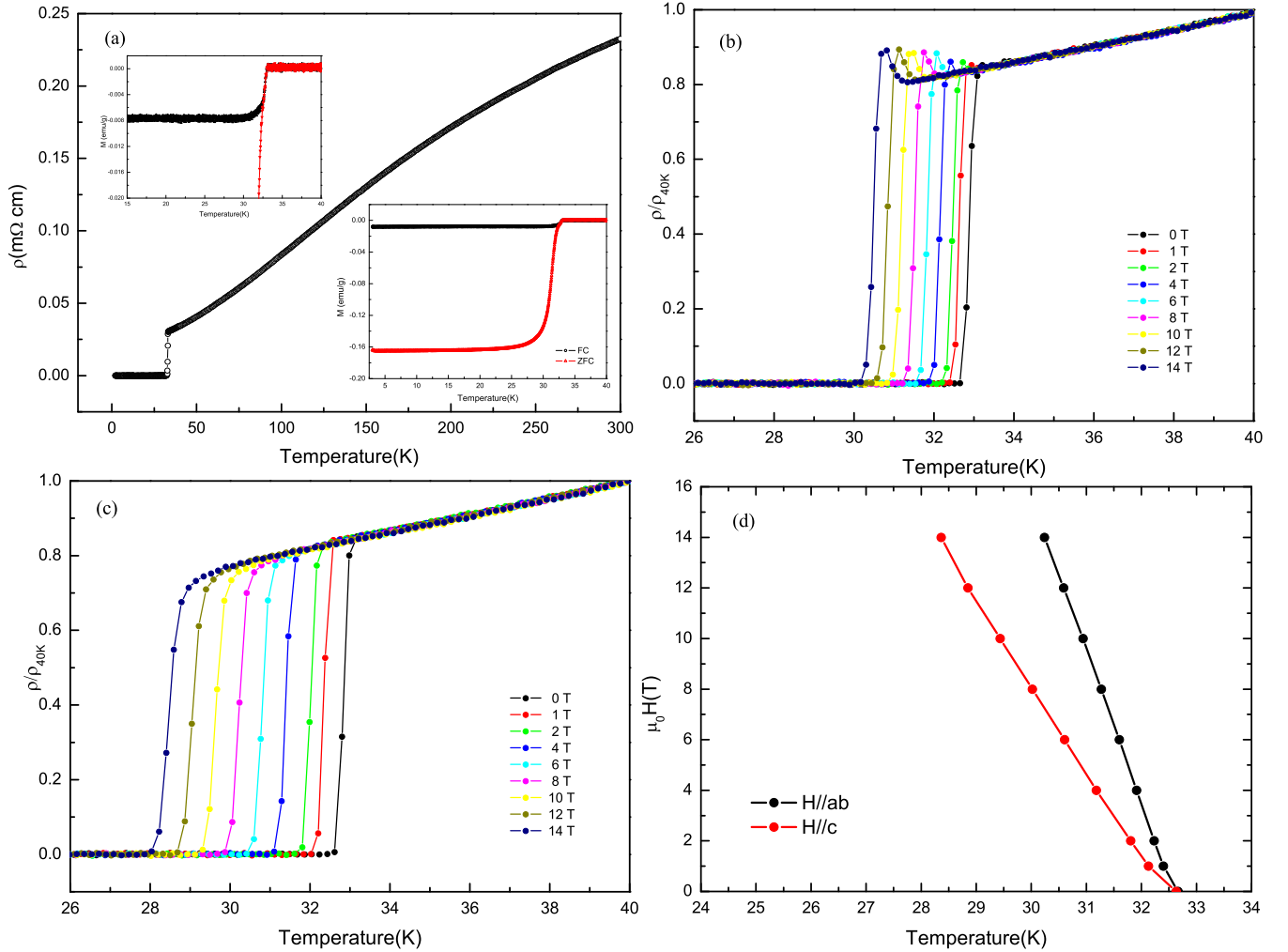}
\caption{(Color online)  (a) Temperature dependence of the
electrical resistivity for a NaFeAs$_{0.8}$P$_{0.2}$ single
crystal in zero field up to 300 K. The down right inset shows the
temperature dependent magnetization data under ZFC and FC
conditions with H=10 Oe. The up left inset is the enlarged part
for M-T curve. (b) Temperature dependence of the resistivity
normalized to resistivity at 40 K at fields up to 14 T with
H$\|$ab. (c) Temperature dependence of the resistivity normalized
to resistivity at 40 K at fields up to 14 T with H$\|$c. (d) The
upper critical field of NaFeAs$_{0.8}$P$_{0.2}$ single crystal for
H$\|$ab and H$\|$c.}
\end{figure}

The above results on electronic properties in Co-doped NaFeAs are
consistent with that of Co-doped 122 or 1111 systems. We know that
phosphor doping in BaFe$_{2}$As$_{2}$ induces supercnductivity
with Tc=30 K, much higher than the Tc=22K of Co-doped
BaFe$_{2}$As$_{2}$. When we dope NaFeAs with P, our results do
show clear transition temperature high up to 33 K. It is clearly
shown both in the resistivity curve (Fig. 2(a)) and magnetization
data (The inset of Fig. 2(a)). The superconducting transition is
very sharp with the transition width smaller than 1 K. What is the
cause of such a high transition temperature remains an open
question. Mizuguhci \emph{et al.} proposed in their report that
the anion height is crucial in determining the Tc of the
iron-based superconductors.\cite{mizuguchi} In P-doped NaFeAs, the
Fe has the valence of $2+$, and no nominal carriers are introduced
into the iron layer, which may not cause disorder in the iron
layer. But according to their strict requirement, no dopings can
be done on anion site either, ie. no substitution of P for As
should be done if the rules are satisfied before P-doped NaFeAs is
put on the symmetric curve they proposed.

On the other hand, regular tetrahedron of FeAs$_{4}$ lattices in the
FeAs block are believed to be one key point in achieving higher
Tc\cite{jzhao,lee}. More and more results are consistent with the
prediction and obey the rule. In 111 system, pressure effect on
LiFeAs is systematically studied, and Tc decreases with increasing
pressure.\cite{mito} The lattice parameters are summarized based on
Rietveld refinement and it is found that the FeAs$_{4}$ tetrahedron
is highly distorted and the pressure further enhances the distortion
so the Tc decreases. As for NaFeAs, the Tc of NaFeAs is 23
K,\cite{gfchen111} and refinement results show us NaFeAs possesses a
more regular tetrahedron with the basic difference that the radii of
Na is bigger than that of Li and it forms a less distorted
environment.\cite{parker} The P-doped NaFeAs superconducts at 33 K,
so we believe that the FeAs$_{4}$ tetrahedron here is more regular
than that in Na-deficiency polycrystalline NaFeAs.\cite{parker} We
should emphasize the Tc is the highest in all Co-doped or
isoelectronic element-doped systems. In previous report on NaFeAs
pollycrystalline samples, the maximum Tc of 31 K is observed when
the pressure of 3.0 GPa is applied.\cite{zhang} As we know the
pressure effect on iron-based superconductors are complex with two
major categories: one type is that pressure enhances the Tc of the
samples while another is that the pressure firstly enhances then
suppresses the Tc. These behaviors are intimately related to the
lattice parameters and configuration of FeAs$_{4}$ tetrahedron.
Considering such effect, we expect possible higher Tc to be realized
in the work studying the pressure effect on P-doped NaFeAs crystals.
Further work on the lattice parameters and pressure effect on the
P-doped NaFeAs are in progress.

Anisotropy of electronic transport properties have also been
studied and the results are shown in Fig.2 (b) and (c). The
similar behavior of field on the transition is observed. When the
magnetic field is applied within ab plane, the transition is less
affected than the situation when the field is applied along the c
axis. The corresponding values are $-dH_{c2}/dT|_{T_{c}}$=6.0 T/K
for $H\|ab$ and $-dH_{c2}/dT|_{T_{c}}$=3.4 T/K for $H\|c$. Taking
33 K as Tc and using WHH formula, we get the corresponding upper
critical fields as high as $H_{c2}^{ab}$=137 T and $H_{c2}^{c}$=77
T. Such high critical fields and low anisotropy ratio
(137/77=1.78) are comparable to other 122 system and may have the
potential for future applications in the superconducting devices.
Recently, Chong \emph{et al.} summarized the results of high field
study on 122 and 111 samples and concluded that the hole-doped
system have higher upper critical fields $>$ 100 T compared to the
electron-doped system, while those of strain or pressure induced
superconductvity have similar upper critical fields as in the
electron-doped system.\cite{chong} They did the measurements on
single crystalline BaFe$_{2}$As$_{1.36}$P$_{0.64}$ (Tc=31 K) and
estimated the upper critical fields as H$_{c2}^{ab}$=77 T and
H$_{c2}^{c}$=36 T. Our results of upper critical fields with
$H_{c2}^{ab}$=137 T and $H_{c2}^{c}$=77 T are the largest among
the 111 and 122 systems except for those in hole-doped cases and
it is the first exception of samples with upper critical field
$>$100 T which is non-hole-doped system. Previous high magnetic
field experiments on (Ba,K)Fe$_{2}$As$_{2}$ show nearly isotropic
superconductivity in that system\cite{hqyuan} and later the same
conclusion is drawn by other high magnetic field experiments on
Ba(Fe,Co)$_{2}$As$_{2}$.\cite{tanatar,yamamoto} So, it is
necessary to do the high magnetic field experiments in the 111
system and the work is in progress.

To summarize, we have grown the high quality single crystals of
NaFeAs with Co- and P-doping. We find that the superconductivity
occurs at ambient pressure with Tc up to 33 K in
NaFeAs$_{0.8}$P$_{0.2}$. We emphasized that the value of Tc is the
highest in 111 system and in all other isoelectronic element
substituted 122 or 1111 system. The upper critical fields are
estimated high up to 77 T with H$\|$c and 137 T with H$\|$ab for
NaFeAs$_{0.8}$P$_{0.2}$, which are also the highest in 122 and 111
systems except for those in hole-doped cases. We believe that
these results provide important information for establishing a
theory to understand the superconductivity in the iron-based
superconductor.

The work is supported by the National Science Foundation of China
and the 973 project of the Ministry of Science and Technology of
China.
%\begin{figure}
%\includegraphics[width=6 cm]{fig2.eps}
%\caption{(Color online) Displacement patterns for the Raman-active modes
%of $\alpha$-FeTe, obtained from LDA calculations. (a,b) E$_{g}$ modes,
%(c) A$_{1g}$ mode of Te atoms, (d) B$_{1g}$ mode of Fe atoms. }
%\end{figure}

\end{document}